\documentclass[10pt,conference]{IEEEtran}
\IEEEoverridecommandlockouts
\usepackage{cite}
\usepackage[hidelinks]{hyperref}
\usepackage{amsmath,amssymb,amsfonts}
\usepackage{algorithmic}
\usepackage{graphicx}
\usepackage{textcomp}
\usepackage{xcolor}
\usepackage{booktabs}
\usepackage{tabularx}
\usepackage{array}

\usepackage[table]{xcolor}

\definecolor{KnowledgeBlue}{HTML}{2D6E95}
\definecolor{SkillsGreen}{HTML}{198176}
\definecolor{RowShade}{HTML}{F1F5F7}
\definecolor{MissingGray}{HTML}{ABB5BD}

\usepackage[most]{tcolorbox}

\newtcolorbox{implicationbox}[1]{
    enhanced,
    breakable,
    colback=black!3,
    colframe=black!45,
    colbacktitle=black!8,
    coltitle=black,
    title={#1},
    fonttitle=\bfseries\small,
    fontupper=\small,
    boxrule=0.4pt,
    arc=1mm,
    left=2mm,
    right=2mm,
    top=1.5mm,
    bottom=1.5mm,
    before skip=8pt,
    after skip=8pt
}

\author{\IEEEauthorblockN{1\textsuperscript{st} R.D.N. Shakya}
\IEEEauthorblockA{
\textit{University of Moratuwa}\\
Katubedda, Sri Lanka \ \\
shakyardn.26@uom.lk}
\and
\IEEEauthorblockN{2\textsuperscript{nd} C.P. Wijesiriwardana}
\IEEEauthorblockA{
\textit{University of Moratuwa}\\
Katubedda, Sri Lanka \ \\
chaman@uom.lk}
\and
\IEEEauthorblockN{3\textsuperscript{rd} S.M.Vidanagamachchi}
\IEEEauthorblockA{
\textit{University of Ruhuna}\\
Matara, Sri Lanka \\
smv@dcs.ruh.ac.lk}
\and

\IEEEauthorblockN{4\textsuperscript{th} Nalin A.G. Arachchilage}
\IEEEauthorblockA{
\textit{RMIT University}\\
Melbourne, Australia \\
nalin.arachchilage@rmit.edu.au}

}

\begin{document}

\title{Post-Quantum Cryptography Education for Software Developers: A Multivocal Literature Review}

\maketitle

\begin{abstract}
Quantum computing threatens widely deployed public-key cryptography, making post-quantum cryptography (PQC) adoption a software-engineering challenge. However, it remains unclear how well existing educational provision prepares software developers for secure PQC implementation and migration. We conducted a multivocal literature review of 8 research publications and 15 professional training offerings, examining competency coverage, practical software-engineering tasks, pedagogical approaches, learner support, and evaluation. All 23 sources covered PQC foundations and algorithms, but coverage of practical software-engineering tasks was uneven. Our analysis identifies six gaps concerning developer-focused evaluation, authentic software-engineering tasks, task–competency alignment, pedagogical design and learner support, security-focused assessment, and long-term validation of learning outcomes and transfer. We derive recommendations for strengthening PQC education and better preparing software developers to implement and migrate cryptographic systems securely.
\end{abstract}


\section{Introduction}
\label{Intro}

Quantum computing threatens widely deployed public-key cryptography. A sufficiently capable quantum computer could undermine the mathematical assumptions underlying cryptographic mechanisms used for digital signatures, authentication, and key establishment \cite{bernstein_post-quantum_2017, dam_survey_2023, BASERI2026104917}.  In response, post-quantum cryptography (PQC) develops cryptographic algorithms designed to resist attacks by both classical and quantum computers \cite{bernstein_post-quantum_2017, dam_survey_2023}. However, transitioning to PQC requires more than replacing vulnerable algorithms; developers must also select, integrate, and migrate cryptographic mechanisms securely across existing and new software systems.

PQC adoption is therefore not solely an algorithm-selection problem. Standards, migration guidance, libraries, and implementation tools facilitate deployment \cite{nist_pqc, bavdekar_post_2023, naether_migrating_2024, hekkala_implementing_2023, ahmed_survey_2025}, but security also depends on how developers configure, integrate, test, and maintain cryptographic functionality \cite{crptoImmerr, bagirovs_applications_2024, toruan2026security}. Incorrect parameter choices, API misuse, inadequate key management, and implementation-level leakage can compromise systems even when the underlying algorithms are sound \cite{crptoImmerr, howe_sok_2021, hekkala_implementing_2023}. 

Developers must therefore identify vulnerable cryptographic dependencies, select appropriate mechanisms, integrate PQC into software and protocols, and validate functionality and security \cite{decisionAlgo, naether_migrating_2024, hekkala_implementing_2023}. These activities demand competencies beyond conceptual awareness of quantum threats and PQC. Previous research reports knowledge and readiness challenges among stakeholders involved in migration \cite{giron_migrating_2023}, while the expertise required varies between developers integrating cryptographic libraries and specialists modifying implementations \cite{decisionAlgo}. Educational preparation must therefore connect PQC foundations with the practical tasks developers are expected to perform.

Existing PQC educational interventions include structured courses, laboratories, and interactive learning environments \cite{R3, R4, GL1}. These approaches must balance cryptographic foundations with practical activities while accommodating learners with different technical backgrounds \cite{R4, parakh_quintet_2024}. However, including practical activities and covering relevant topics do not necessarily demonstrate secure practical competence. Conceptual tests and perception-based measures alone may be insufficient; assessment should also examine the security of learners’ implementations \cite{practComAssess}. Constructive alignment emphasises consistency between learning outcomes, activities, and assessment \cite{biggs1996alignment}, while valid performance assessment requires relevant tasks and explicit criteria \cite{nrc2001assessment}. In cryptographic programming, functional correctness does not guarantee secure API usage \cite{acar2017cryptographic}, and performance on familiar exercises does not establish transfer to unfamiliar contexts \cite{nrc2000learning}.

Although existing PQC reviews examine algorithms, tools, implementation barriers, and migration processes \cite{shakya_sok_pqc_implementation_2026, ahmed_survey_2025, naether_migrating_2024}, evidence on developer education remains distributed across research papers and professional training offerings (hereafter, training offerings). We address this gap through a \emph{multivocal literature review} (MLR) of eight research papers and 15 training offerings. We characterise learning approaches and target audiences, map documented knowledge, skills, and developer tasks against the NICE Workforce Framework for Cybersecurity \cite{NICER1}, and examine pedagogical support and evidence of learner capability. We address four research questions:

\begin{itemize}
\item \textbf{RQ1:} What learning approaches are available for PQC education, and to what extent do they target software developers?
\item \textbf{RQ2:} What PQC knowledge, skills, and practical software-engineering tasks do these approaches address, and to what extent are tasks explicitly aligned with competencies?
\item \textbf{RQ3:} What pedagogical strategies, interaction mechanisms, and forms of learner support do these approaches provide?
\item \textbf{RQ4:} How are these approaches assessed and evaluated, and what evidence demonstrates their contribution to secure and transferable PQC implementation and migration capabilities?
\end{itemize}

This study makes three contributions. First, it provides a comparative evidence map of research papers and training offerings, covering target learners, practical tasks, pedagogy, and assessment. Second, it distinguishes stated competencies and reviewer-derived NICE mappings from explicit task-competency alignment and demonstrated learner performance. Third, it identifies evidence-informed priorities for authentic developer tasks, security-focused assessment, and evaluating retention and transfer.

\section{Related Work}
\label{RelatedWork}

Research on PQC adoption has examined technical and socio-technical challenges, cryptographic library support, and migration challenges \cite{shakya_sok_pqc_implementation_2026, ahmed_survey_2025, naether_migrating_2024}. Although these studies characterise challenges in PQC implementation and migration, they do not systematically examine whether existing educational approaches prepare software developers to address them. A complementary body of research has examined how cryptographic knowledge and skills are taught, focusing on instructional design and learner activities in coding tutorials and cryptography education \cite{OnlineCourseAnalysis, CryptEduReview}. Although PQC-specific studies have introduced courses and laboratory environments \cite{R4, R3}, their findings have not been synthesised alongside publicly available training offerings in terms of authentic developer tasks, competency alignment, pedagogical support, and assessment validity. Our review addresses this gap by mapping PQC educational provision and evaluation for software developers.

\subsection{NICE Framework}
\label{NICEFramework}

The NICE Workforce Framework for Cybersecurity (NIST SP 800-181 Rev. 1) provides a common vocabulary for describing cybersecurity work through task, knowledge, and skill statements \cite{NICER1}. It can support the analysis and design of cybersecurity education by relating the capabilities learners need to develop to the tasks they are expected to perform. In this study, we use the NICE Framework as an analytical reference for classifying PQC-related knowledge and skills reported in educational interventions and training offerings. We distinguish reviewer-derived competency mappings from provider-declared competency specifications, explicit task-competency alignment, and evidence of assessed learner capability. This distinction enables us to examine not only which competencies educational approaches appear to address, but also whether the sources explicitly connect those competencies to practical software-engineering tasks and assess their application.

\section{Methodology}
\label{Method}

\begin{figure*}[t]
\centering
\includegraphics[width=0.7\textwidth,height=0.45\textheight]{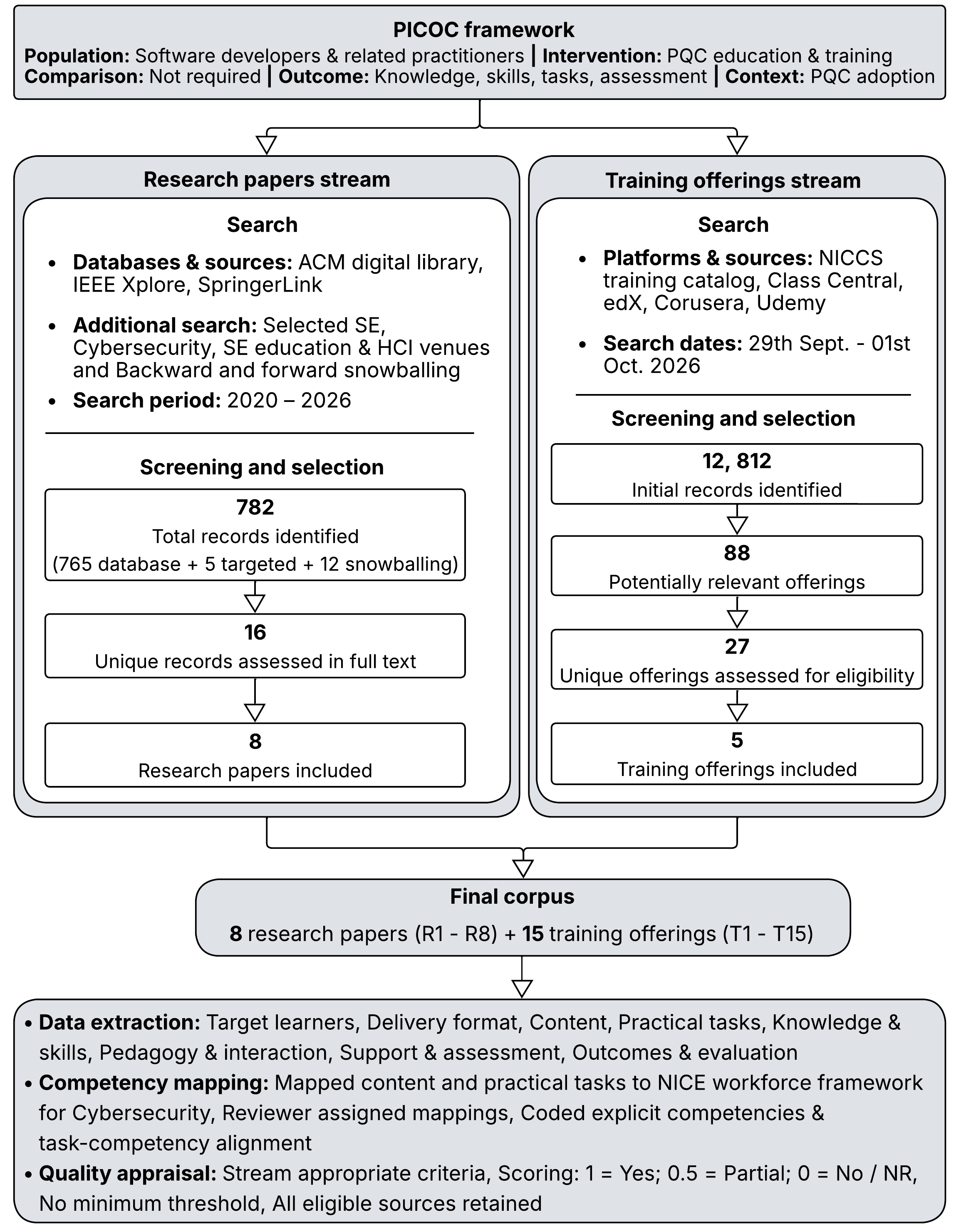}
\caption{Multivocal review and evidence-mapping procedure}
\label{MethodFigure}
\end{figure*}

We conducted a multivocal literature review (MLR) combining peer-reviewed research papers and publicly available training offerings to examine PQC education, its content, and its evaluation. The research stream followed structured review procedures informed by Kitchenham and Charters \cite{kitchenham_procedures_2004}, while the training offerings stream followed Garousi et al.'s MLR guidelines \cite{MLR}. We searched, screened, and extracted data separately for each stream before synthesising the findings. The final corpus comprised eight research papers (R1-R8) and 15 training offerings (T1-T15). Figure \ref{MethodFigure} summarises the workflow.

\subsection{Review Scope and Search Strategy}
\label{ScopeSearch}

We used the Population, Intervention, Comparison, Outcome, and Context (PICOC) framework to scope the review. The population included software developers and related technical practitioners, including software engineers, cybersecurity and IT professionals, and computing learners. Interventions covered PQC education, training, practical activities, and learner support. Outcomes concerned knowledge, skills, software engineering tasks, assessment, and evidence of learning and capability in the context of PQC adoption. We did not require a comparator because the aim was to map educational approaches rather than estimate comparative effects.

For the research stream, we searched ACM Digital Library, IEEE Xplore, and SpringerLink in August 2026 for publications from 2020 to 2026. Search terms combined software developers and related practitioners, education and training, and PQC variants, including post-quantum, quantum-resistant, and quantum-secure cryptography. We adapted queries to each database's syntax and length restrictions. We excluded outcome terms to capture educational approaches regardless of whether they reported evaluations. We also searched selected software engineering, cybersecurity, software engineering education, and human-computer interaction venues, followed by backward and forward snowballing.

For the training offerings stream, we searched the NICCS training catalogue \cite{niccsCat}, Class Central \cite{classcentral}, edX \cite{edxplatform}, Coursera \cite{courseraplatform}, and Udemy \cite{udemyplatform} on 29 September and 1 October 2026. Search terms included ``post-quantum cryptography'', ``post quantum cryptography'', ``post-quantum'', and ``quantum-safe cryptography''. We screened titles and available metadata for substantive PQC relevance, excluding offerings focused on quantum computing without relevant PQC learning. We also applied a predefined duration filter, excluding video-based offerings of 30 minutes or less.

\subsection{Eligibility and Selection}
\label{EligSelec}

For the research stream, we included peer-reviewed journal and conference publications published between 2020 and 2026 that were written in English, available in full text, and described educational, training, learning, or developer-support approaches relevant to PQC adoption among software developers or related practitioners. We excluded non-peer-reviewed papers, sources without an educational component, inaccessible full texts, and non-English papers.

For the training offerings stream, we included publicly documented courses, programmes, modules, workshops, and other structured interventions addressing PQC or quantum-safe cryptography as a dedicated or substantial component. Offerings had to target or apply to relevant technical learners and provide sufficient information to address at least one research question. We excluded unstructured offerings, incidental PQC mentions, quantum computing-focused offerings without substantive PQC relevance, offerings targeting exclusively non-technical audiences, insufficiently documented or inaccessible sources, and duplicates. Video-based offerings of 30 minutes or less were also excluded.

For the research stream, database searches identified 765 records. Preliminary screening and duplicate checking left 16 unique records for full-text assessment. Targeted searches identified five records but yielded no additional eligible papers. Snowballing identified 12 records, of which five passed preliminary screening; all five were duplicates. Full-text screening retained eight papers.

For the training offerings stream, searches identified 12,812 records. Title and metadata screening retained 88 potentially relevant offerings. After duplicate removal, 27 unique offerings underwent eligibility assessment, and 15 met the inclusion criteria. We documented selection using a PRISMA-inspired reporting structure \cite{page_prisma_2021}. The final corpus contained eight research papers and 15 training offerings.

\subsection{Data Extraction and Synthesis}
\label{Data_Extraction}

We extracted data on target learners, delivery formats, conceptual content, practical tasks, knowledge and skills, pedagogy, interaction, learner support, assessment, reported outcomes, and empirical evaluation. A structured codebook was applied to both streams, allowing multiple codes per source where supported by evidence. Descriptive counts and cross-source matrices summarised educational features, while qualitative synthesis examined relationships between learning outcomes, activities, practical tasks, and evidence of capability, informed by thematic-analysis principles \cite{braun_using_2006}.

We classified evidence as explicitly present, explicitly absent, or not reported (NR), where supported by the source. Missing evaluation information in training offerings descriptions was coded as NR, not as proof that evaluation had not occurred. We distinguished the provision of practical tasks from the assessment of learner performance, and task completion from meeting explicit security and correctness criteria.

To assess coding consistency, reviewers independently compared coding decisions for competency and content categories, obtaining an initial percentage agreement of 75.603\%. They discussed and resolved disagreements to reach consensus on the final coding.

\subsection{Competency Mapping and Quality Appraisal}
\label{StudyComMapping}

For RQ2, we mapped documented learning content and practical activities to relevant knowledge and skill statements in the selected version of the NICE Workforce Framework for Cybersecurity \cite{NICER1}. These reviewer-assigned mappings indicate correspondence with competency statements, not provider-declared alignment or demonstrated learner mastery. We separately coded whether sources explicitly specified competencies and linked individual tasks to them.

We appraised research papers and training offerings using stream-appropriate criteria informed by Garousi et al. \cite{MLR}. Each criterion received a score of 1 (yes/adequate), 0.5 (partially), or 0 (no or not reported), with scores converted to percentages. We applied no minimum quality threshold; all eligible sources were retained, and appraisal findings were used to contextualise the synthesis, consistent with guidance that appraisal need not determine inclusion \cite{qaThreashold}. Scores were interpreted separately for each stream because their purposes and reporting conventions differ. We considered reporting limitations when assessing the verifiability, comparability, and reproducibility of documented educational provision, without assuming that unreported activities or outcomes were absent.

\section{Results}
\label{Results}

The final evidence set comprised 8 research papers (R1–R8) and 15 training offerings (T1–T15). Each source was assigned a unique identifier to facilitate traceability between the extracted data, synthesized findings, and underlying evidence. Table \ref{sourceIndex} maps these identifiers to their corresponding bibliographic references. We use these identifiers throughout the results to indicate which sources support each reported observation. We analysed the sources against the four research questions, distinguishing documented educational provision from empirical evidence of learning outcomes. The findings are summarised through six evidence-based observations (F1–F6). 

\subsection{Quality Appraisal Results}
\label{QAResults}

Training offerings had a mean appraisal score of 93.89\% (median: 91.67\%; range: 87.50-100\%), compared with 77.08\% for research papers (median: 83.33\%; range: 50.00-91.67\%). These scores are interpreted separately because each stream used different appraisal criteria. Five of eight research papers partially met the data-analysis criterion, and all eight partially met the transparency and reproducibility criterion. Eight of 15 training offerings lacked an identifiable publication, update, or current-status date. These limitations informed interpretation of the synthesis but did not warrant excluding eligible sources or treating unreported activities and outcomes as absent.

\begin{table}[t]
\caption{Included research papers and training offerings sources with corresponding bibliographic references}
\label{sourceIndex}

\begin{tabularx}{\columnwidth}{@{}lXl@{}}
\toprule

\textbf{ID} & \textbf{Source title} & \textbf{Ref.} \\
\midrule

\multicolumn{3}{@{}l}{\textit{Research papers}} \\

R1 & PQCIP: A Post-Quantum Cryptography Educational Program for Cybersecurity Professionals & \cite{R1} \\

R2 & A Lightweight Plug-in Module for Introducing Post-Quantum Cryptography in Higher Education: An Experience Report & \cite{R2} \\

R3 & Bridging Theory and Practice via PQC Hands-On Laboratories in Cybersecurity Education & \cite{R3} \\
  
R4 & Towards a Quantum-Resistant Future: Experiences in Post-Quantum Cryptography Education
& \cite{R4} \\

R5 & Modular Approach to Teaching Post-Quantum Cryptography & \cite{R5} \\

R6 & Designing and Delivering a Post-Quantum Cryptography Course & \cite{R6} \\

R7 & Quantum-Ready Education: Preparing Students for Post-Quantum Computing Societies & \cite{R7} \\

R8 & An Interactive Tool for Teaching Lattice-Based Post-Quantum Cryptography & \cite{R8} \\

\midrule 

\multicolumn{3}{@{}l}{\textit{Training offerings}} \\

T1 & Certified Post-Quantum Cybersecurity Expert (CPQCE) & \cite{GL1} \\

T2 & Certified Quantum Cryptography Architect (CQCA) & \cite{GL2} \\

T3 & Introduction to Quantum Computing Security for Programmers (IQCSP) & \cite{GL3} \\

T4 & Certified Quantum \& Post-Quantum Cryptography Professional (QPQCP) & \cite{GL4} \\

T5 & Quantum-Resistant Cryptography and Its Defense Applications Training & \cite{GL5} \\
 
T6 & Post-Quantum Authentication and Key Infrastructure (PQAKI) Essentials & \cite{GL6} \\

T7 & Certified Quantum Cryptography and Security Specialist (CQCSS) & \cite{GL7} \\

T8 & Introduction to Post-Quantum Cryptography & \cite{GL8} \\

T9 & Quantum Computing-Risks to Encryption and the Implications & \cite{GL9} \\

T10 & Post-Quantum Cryptography Foundations & \cite{GL10} \\

T11 & The Quantum Computing Risk and Security Masterclass 2026 & \cite{GL11} \\

T12 & Introduction to Post-Quantum Cryptography & \cite{GL12} \\

T13 & Post-Quantum Crypto Migration: Inventory, Plan, Execute & \cite{GL13} \\

T14 & Post-Quantum Cryptography: Applied PQC \& NIST Standards & \cite{GL14} \\

T15 & Post-Quantum Cryptography: The NIST Standards Explained & \cite{GL15} \\

\bottomrule
\end{tabularx}
\end{table}

\subsection{RQ1: Learning Approaches and Target Populations}

\textbf{F1: Academic interventions primarily address student populations, whereas training offerings target broader occupational groups}

As shown in Table \ref{SourceCharac}, seven research papers involved university or high-school learners (R2–R8), whereas R1 targeted cybersecurity and IT professionals alongside academic participants. None explicitly identified software developers as a distinct target population. By comparison, 14 professional courses targeted cybersecurity professionals (T1–T11, T13–T15), ten included IT professionals (T4–T6, T8–T11, T13–T15), and five explicitly identified software developers (T3, T8, T9, T14, T15).

Course-based instruction was prevalent across both evidence streams, appearing in four research papers (R1, R4–R6) and 14 training offerings (T1–T7, T9–T15). Alternative approaches included module (R2), laboratory-based learning (R3), a multi-component intervention (R7), an interactive educational tool (R8), and a workshop (T8). These approaches differed in delivery mechanisms, practical activities, and intended learner backgrounds.

The results indicate that PQC education has been investigated primarily in academic populations, whereas publicly available training offerings address a wider occupational audience. However, explicitly developer-oriented provision remains limited across the reviewed sources.

\begin{table}[t]
\centering
\caption{Characteristics of the included sources}
\label{SourceCharac}

\begin{tabularx}{\columnwidth}{>{\raggedright\arraybackslash}Xrr}
\toprule
\textbf{Characteristic} & \textbf{Research papers} & \textbf{Training offerings} \\
\midrule
Student learners & 7/8 & 1/15 \\
Cybersecurity professionals & 1/8 & 14/15 \\
IT professionals & 1/8 & 10/15 \\
IT/risk managers & 0/8 & 9/15 \\
Explicitly developer-targeted & 0/8 & 5/15 \\
Course based & 4/8 & 14/15 \\
Empirical evaluation reported & 6/8 & 0/15 \\
\bottomrule
\end{tabularx}
\end{table}

\subsection{RQ2: Knowledge, Skills, and Practical Software-Engineering Tasks}

\textbf{F2a: Foundational PQC topics were widely documented, while adoption-related topics received uneven coverage}

\begin{figure*}[t]
    \centering
    \includegraphics[width=0.9\textwidth]{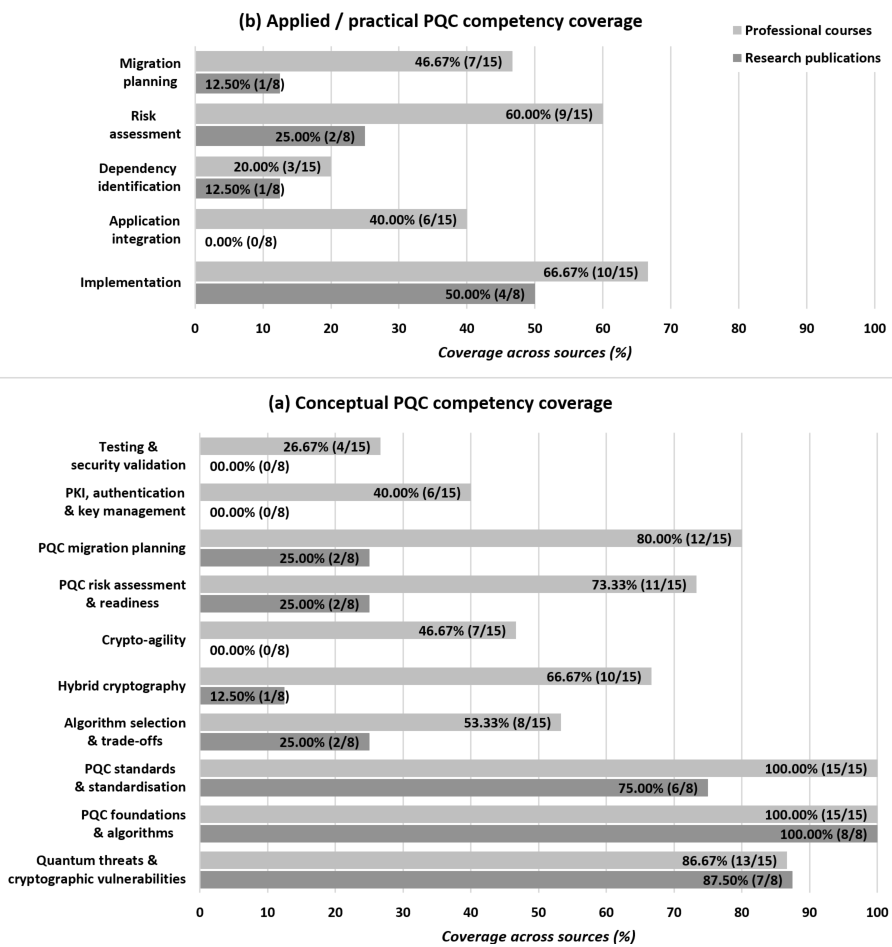}
    \caption{Documented conceptual PQC topic coverage (a) and documented applied/practical PQC task or capability coverage (b) by source type. Bar lengths indicate the percentage of sources within each group, and labels show source counts. Categories are non-exclusive. Practical classifications indicate documented source-level task or capability coverage, not evidence of learner performance or competency acquisition.}
    \label{contentCoverage}
\end{figure*}

As shown in Fig. \ref{contentCoverage}(a), all 23 sources documented coverage of PQC foundations and algorithms (R1-R8; T1-T15). Twenty sources documented quantum threats and cryptographic vulnerabilities (R1-R7; T1-T13), while 21 sources documented PQC standards and standardisation (R1, R2, R4-R7; T1-T15).

Coverage was less consistent for topics associated with operational PQC adoption. Algorithm selection and trade-offs were documented in two research papers (R1, R7) and eight training offerings (T2, T4, T5, T8-T10, T14, T15). Hybrid cryptography was documented in one research paper (R5) and ten training offerings (T2-T4, T6, T7, T9, T10, T13-T15). Crypto-agility was not identified in the research papers but was documented in seven training offerings (T2, T4, T6, T9, T11, T13, T15). Risk assessment and readiness were documented in 2 research papers (R1, R7) and 11 training offerings (T1, T2, T4-T11, T13). PKI, authentication, and key management; and testing and security validation were documented only in training offerings.

Overall, the sources shared a foundation in PQC concepts and algorithms, whereas training offerings more frequently documented topics associated with adoption, infrastructure, and operational considerations. These findings describe the distribution of documented content; they do not, by themselves, establish the depth, instructional effectiveness, or learning outcomes associated with that content.

\textbf{F2b: Documented applied/practical software engineering task coverage differed between research papers and training offerings}

As shown in Fig. \ref{contentCoverage}(b), implementation was documented as an applied/practical task or capability in four research papers (R1, R5-R7) and 10 training offerings (T3, T5, T7, T9, T10, T12-T15). Application integration was documented at this level in 6 training offerings (T5, T6, T9, T10, T13, T15), but in none of the eight research papers. Other adoption related tasks were documented in fewer research papers: dependency identification appeared in one (R2), compared with 3 training offerings (T9, T13, T14); risk assessment appeared in two (R1, R7), compared with 10 training offerings (T1, T2, T4, T6, T7, T9-T11, T13, T14); and migration planning appeared in one (R1), compared with seventraining offerings (T4, T8, T10, T11, T13-T15).

The distinction between conceptual coverage and documented applied/practical tasks was particularly evident for migration planning. Although 12 training offerings (T2, T4, T6, T7-T15) covered migration planning conceptually, only seven (T4, T8, T10, T11, T13-T15) met the coding criterion for documenting it as an applied/practical task or capability. 

A source was coded as documenting an applied/practical PQC competency only when its available description specified a task-oriented activity, practical exercise, implementation procedure, or intended practical capability. Topic mentions and conceptual descriptions alone were insufficient. These codes represent documented educational coverage, not verified learner performance, assessment, or competency acquisition. 

\textbf{F3: Intended competencies are commonly specified, but explicit task–competency alignment remains undocumented}

Table \ref{ComPedSupEva} shows that 20 of the 23 sources specified intended competencies or learning capabilities (R1, R3, R5–R7; T1–T15). Nevertheless, none explicitly reported systematic alignment between individual developer tasks and the knowledge or skills required to perform them under the review's coding criteria. All task–competency alignment entries were coded as not reported (NR), rather than as confirmed absence. For example, T9, T13, and T14 described application integration and migration activities without providing explicit task-level competency mappings.

The reviewer-derived NICE Framework mapping identified correspondences between the reported educational content and cybersecurity knowledge and skill statements. However, these mappings represent analytical classifications performed during the review rather than evidence that the original interventions employed NICE-based instructional design. Although the source content could be mapped to NICE Framework knowledge and skill statements, these reviewer-derived mappings do not demonstrate provider-declared competency alignment or successful acquisition of the associated competencies.

\begin{table}[t]
\centering
\caption{Documented evidence of competency specification, pedagogy, learner support, assessment, and evaluation}
\label{ComPedSupEva}
\begin{tabularx}{\columnwidth}{
    >{\raggedright\arraybackslash}X
    >{\centering\arraybackslash}p{0.82cm}
    >{\centering\arraybackslash}p{1.25cm}
}
\toprule
\textbf{Indicator} & \textbf{Research papers} & \textbf{Training offerings} \\
\midrule

\multicolumn{3}{@{}l}{\textit{Competency specification and alignment}} \\
Explicit competency specification & 5/8 & 15/15 \\
Explicit task--competency alignment & 0/8 & 0/15 \\
\midrule

\multicolumn{3}{@{}l}{\textit{Pedagogical approaches}} \\
Lecture-based instruction & 6/8 & 10/15 \\
Problem-based learning & 1/8 & 0/15 \\
Project-based learning & 1/8 & 7/15 \\
\midrule

\multicolumn{3}{@{}l}{\textit{Learning activities and practical engagement}} \\
PQC implementation or integration coding tasks & 3/8 & 3/15 \\
Interaction with tools & 6/8 & 2/15 \\
\midrule

\multicolumn{3}{@{}l}{\textit{Learner support}} \\
Explicit feedback mechanisms & 3/8 & 0/15 \\
Additional learning resources & 2/8 & 11/15 \\
Scaffolding & 1/8 & 0/15 \\
Adaptive or personalised support & 0/8 & 0/15 \\
\midrule

\multicolumn{3}{@{}l}{\textit{Assessment methods}} \\
Quizzes & 2/8 & 6/15 \\
PQC implementation or integration coding tasks & 1/8 & 1/15 \\
Projects & 1/8 & 4/15 \\
Laboratory practical exercises & 3/8 & 0/15 \\
\midrule

\multicolumn{3}{@{}l}{\textit{Assessment criteria and evidence}} \\
Knowledge assessed & 3/8 & 3/15 \\
Practical performance assessed & 3/8 & 1/15 \\
Security criteria explicitly reported & 2/8 & 0/15 \\
\midrule

\multicolumn{3}{@{}l}{\textit{Evaluation outcomes}} \\
Empirical evaluation reported & 6/8 & 0/15 \\
Pre/post knowledge outcomes reported & 4/8 & 0/15 \\
Learner perceptions or usability reported & 3/8 & 0/15 \\
Delayed knowledge retention established & 0/8 & 0/15 \\
Transfer of learning demonstrated & 0/8 & 0/15 \\
\bottomrule
\end{tabularx}

\vspace{2pt}
\parbox{\columnwidth}{\scriptsize
Counts indicate the number of sources explicitly reporting each feature out of the total sources in each category. A zero indicates no explicit report under the adopted coding criteria and does not necessarily establish the absence of the feature. ``No'' and ``NR'' remain distinct in the underlying extraction.}
\end{table}

\subsection{RQ3: Pedagogical Strategies, Interaction Mechanisms, and Learner Support}

\textbf{F4: Existing approaches incorporate interactive learning mechanisms and practical activities, but structured learner support and adaptation are insufficiently documented}

As summarised in Table \ref{ComPedSupEva}, lecture-based instruction appeared in six research papers (R1–R3, R6–R8) and ten training offerings (T1, T2, T4, T7, T9–T11, T13–T15). Other interventions incorporated tool interaction (R1, R3, R4, R6–R8), simulation (R3, R7, R8), and laboratory activities (R3, R7). R6 included problem-based homework, while R7 used project-oriented learning.

Training offerings similarly described scenario-based activities (T11, T12, T14) and project-based activities (T1, T2, T4, T10, T12–T14). The coding-task indicator in Table \ref{ComPedSupEva} further shows that activities related to PQC implementation or integration were reported in three research papers (R2, R6, R7) and three training offerings (T3, T12, T13). These findings indicate that practical learning activities are present in both evidence streams, although their inclusion alone does not establish whether learners' performance is assessed or whether implementations are evaluated against explicit security criteria.

Learner support was less consistently documented. Explicit feedback mechanisms were reported in three research papers (R2, R3, R8). Additional learning resources were reported in two research papers (R1 and R6) and eleven training offerings (T1, T2, T4, T7, T9–T15). Scaffolding was explicitly reported in only one research paper (R7) and none of the training offerings. No source explicitly documented adaptive or personalised learning mechanisms under the adopted coding criteria.

Overall, the findings indicate that existing approaches combine lecture-based instruction with interactive mechanisms and, in some cases, practical implementation activities. Nevertheless, the available descriptions provide limited evidence of how learner support responds to prior competencies or performance during learning. This distinction is important because offering practical activities is not equivalent to establishing whether learners can perform the tasks successfully and securely.

\subsection{RQ4: Assessment, Evaluation, and Evidence of Secure PQC Capability}

\textbf{F5: Reported evaluations provide limited evidence of security-correct practical performance}

As shown in Table \ref{ComPedSupEva}, six research papers reported empirical educational evaluation (R1, R2, R3, R6, R7, R8), whereas none of the fifteen training offerings explicitly reported empirical learner evaluation in the descriptions examined. Among the research papers, three reported knowledge-related outcomes (R1, R2, R7), four reported pre/post outcomes (R1, R3, R7, R8), and three reported learner perception or usability outcomes (R1, R2, R6). The training offerings' descriptions reported knowledge-related outcomes for three offerings (T9, T10, T12), but did not report pre/post outcomes or learner perception or usability outcomes under the adopted coding criteria.

The reported assessment mechanisms also varied across sources. Research papers described quizzes (R1, R2) and practical laboratory or exercise activities (R1, R3, R7); R7 additionally reported a coding task and a project. Training offerings described quizzes (T1, T2, T4, T7, T9, T12), coding exercise (T3), and projects (T1, T2, T4, T14). These assessment formats indicate that some sources include opportunities to check knowledge or engage with practical work, but their presence alone does not establish that learners' practical performance is evaluated systematically.

Under the adopted coding criteria, practical performance was explicitly assessed in three research papers (R1, R3, R7) and one training offering (T3). Identifiable security-related assessment criteria were reported in two research papers (R3, R7), with none reported in the training offerings. Thus, the evidence becomes more limited when moving from the reporting of assessment activities to explicit evaluation of practical performance and security correctness. In particular, the available descriptions do not establish whether implementation or integration activities consistently assess learners' ability to produce secure PQC implementations.

Taken together, the findings distinguish practical learning activities from assessments of practical performance and evaluations of security correctness. Although activities related to PQC implementation or integration were reported in both evidence streams, only a subset of sources explicitly assessed practical performance, and fewer reported identifiable security-related criteria. Consequently, the available evidence does not establish that these approaches consistently develop demonstrable capability for secure PQC implementation and integration.

\textbf{F6: Evidence of retained and transferable PQC capability remains insufficient}

None of the eight research papers explicitly demonstrated transfer of learning to unfamiliar software-engineering tasks under the adopted coding criteria. The available evidence also did not establish delayed retention conclusively. R3 referred to planned retention assessment, while R7 reported retention scores at a four-week follow-up; however, the available reporting did not clearly establish the delayed assessment conditions sufficiently to classify retained capability conclusively.

None of the fifteen training offerings reported evidence of delayed retention or transfer. Consequently, the evidence remains concentrated on reported learning outcomes and selected assessment activities rather than demonstrated capability retained over time or applied in unfamiliar software-engineering contexts. This limits the conclusions that can be drawn about whether learners retain PQC-related competencies and transfer them successfully to new development tasks.

\section{Discussion}
\label{Discussion}

The findings identify six related gaps in the evidence on preparing software developers for secure PQC adoption. These concern developer-focused evaluation, authentic software-engineering task coverage, task-competency traceability, pedagogical design and learner support, security-focused assessment, and the validation of retained and transferable capability. Together, these gaps highlight a distinction between providing PQC-related educational content and establishing whether learners can apply the required knowledge and skills securely in software-engineering contexts. Table \ref{gap_synthesis} summarises the findings, associated research gaps, and recommendations.

\begin{table*}[t]
\centering
\caption{Evidence-derived findings, research gaps, and recommendations}
\label{gap_synthesis}
\begin{tabularx}{\textwidth}{p{0.1cm}>{\raggedright\arraybackslash}Xp{3.3cm}p{6.4cm}}

\toprule
\textbf{ID} & \textbf{Finding} & \textbf{Research gap }& \textbf{Recommendation} \\
\midrule
F1 & Limited explicit focus on software developers and limited reported professional evaluation
& G1: Developer validity
& Evaluate learning with practising developers and representative engineering tasks \\

F2 & Uneven documentation of applied PQC adoption tasks
& G2: Authentic task coverage
& Integrate realistic dependency discovery, application integration, testing, and migration tasks \\

F3 & Intended competencies are specified, but explicit task--competency alignment is not reported
& G3: Task-competency traceability
& Map tasks to required knowledge, skills, learning outcomes, and assessment criteria \\

F4 & Interactive and practical activities are documented, but adaptive support is not
& G4: Pedagogical design and learner support
& Evaluate scaffolded learning, feedback, and competency-responsive support \\

F5 & Practical performance and security-focused assessment are insufficiently documented
& G5: Security-focused assessment
& Assess functional correctness and security correctness separately \\

F6 & Retention and transfer of learning are not clearly established
& G6: Long-term capability validation
& Conduct delayed assessments and transfer tasks in unfamiliar contexts \\
\bottomrule
\end{tabularx}
\end{table*}

\subsection{G1: Developer Population and Industry Relevance Gap}

\textbf{Interpretation:} F1 identifies a gap between the populations represented in research papers and the intended audiences of training offerings. Although some training offerings target developers, their descriptions do not establish whether they effectively prepare learners for PQC-related engineering work. Similarly, findings from educational interventions involving students cannot automatically be assumed to apply to practising developers, whose work may involve existing architectures, third-party dependencies, cryptographic APIs, interoperability constraints, and legacy systems \cite{acar2017cryptographic}.

The issue is therefore not simply whether PQC education is available to developers, but whether its effectiveness has been evaluated against the demands of professional software development. This highlights the importance of assessing whether learning outcomes demonstrated in educational settings align with the skills required in industry \cite{eduGapIndu}. However, the absence of reported evaluation methods or results in training descriptions should not be interpreted as evidence that providers conduct no evaluation.

\begin{implicationbox}{Recommendation:}
Future studies should evaluate PQC education with practising software developers or participants undertaking representative software engineering tasks. Researchers should document participants' prior experience and its relationship to learning outcomes. Studies involving students should explain the relevance of tasks and evaluation conditions to professional practice. Training providers should report evaluation methods and outcomes clearly to enable assessment of course effectiveness.
\end{implicationbox}

\subsection{G2: Authentic Software Engineering Task Coverage Gap}

\textbf{Interpretation:} F2 indicates that documented coverage of PQC concepts does not necessarily extend to applied software-engineering activities. The difference is important because adoption involves more than understanding algorithms or standards: developers may need to identify cryptographic dependencies, select appropriate algorithms, integrate libraries, test implementations, assess risks, and plan migration \cite{decisionAlgo, naether_migrating_2024, hekkala_implementing_2023}. The reviewed sources vary in the extent to which they document these activities as practical tasks or capabilities.

This distinction is particularly important when interpreting training offerings descriptions. A course may discuss an engineering responsibility conceptually without specifying an activity through which learners practise it. Conversely, documenting a practical activity does not, in itself, demonstrate that learners successfully complete the activity or attain the corresponding competency \cite{biggs1996alignment}.
The observed differences between research papers and training offerings therefore identify variation in documented educational coverage, not differences in educational effectiveness.

The gap has two related dimensions. First, the documented practical treatment of some adoption-related responsibilities is uneven across the evidence streams. Second, the available descriptions do not consistently establish how conceptual content is translated into authentic, observable engineering performance. Addressing both dimensions is necessary to evaluate whether educational approaches prepare learners for the practical demands of PQC adoption.

\begin{implicationbox}{Recommendation:}
Future PQC education should combine conceptual knowledge with authentic software engineering tasks, including dependency discovery, algorithm selection, hybrid deployment, API integration, testing, security validation, and migration planning. Learners should produce practical artefacts, such as code changes, configurations, test results, and migration plans. Assessment should prioritise artefact quality and demonstrated task performance over topic coverage or participation alone.
\end{implicationbox}

\subsection{G3: Task-Competency Alignment Gap}

\textbf{Interpretation:} F3 identifies a traceability gap between intended competencies and the engineering tasks through which those competencies are expected to develop and be demonstrated. Although many sources specify intended competencies or learning capabilities, the reviewed descriptions do not explicitly report systematic alignment between individual tasks and the knowledge and skills required to perform them. All task-competency alignment entries were coded as not reported (NR); this indicates insufficient documentation under the review criteria, rather than confirmed absence of alignment in the underlying instructional designs.

This limitation reduces the transparency of educational design. Without an explicit account of the relationship between tasks, required competencies, learning outcomes, and assessment criteria, it is difficult to determine what learners are expected to apply, how instructional activities support the development of those capabilities, and what evidence would demonstrate successful performance \cite{biggs1996alignment}. It also complicates comparisons between educational approaches because similar objectives may be operationalised and assessed differently.

The reviewer-derived NICE Framework mapping provides an analytical means of relating reported content to cybersecurity knowledge and skill statements \cite{NICER1}. However, these mappings are review-generated classifications, not evidence that the original providers used NICE to design their instruction or that learners acquired the mapped competencies. Moreover, the extent to which existing competency statements adequately represent specialised PQC engineering tasks remains an open question.

\begin{implicationbox}{Recommendation:}
Future PQC education should explicitly link learning tasks to required knowledge, skills, activities, and assessment criteria, using the NICE Framework or other relevant frameworks as references. Researchers should distinguish provider-specified competencies from reviewer-assigned mappings and investigate whether existing frameworks adequately cover specialised PQC tasks. Where gaps exist, they should develop and validate task-specific performance indicators with practitioner input.
\end{implicationbox}

\subsection{G4: Pedagogical Design and Learner Support Gap}

\textbf{Interpretation:} F4 shows that PQC education already incorporates a range of instructional approaches, including interactive tools, simulations, laboratory activities, projects, and implementation-related tasks, suggesting that active learning is already represented in PQC education. However, the documentation of these activities provides less insight into how learners are guided through increasingly demanding tasks or how support responds to their prior knowledge and demonstrated performance.

The distinction is between offering an interactive learning opportunity and establishing how it supports learning. For example, a practical exercise may provide experience with a cryptographic tool, but this alone does not establish whether learners can perform related tasks independently. This transition depends in part on instructional design, including feedback, scaffolding, and the gradual withdrawal of support \cite{feedbck, scafolding}. Prior research suggests that fading support and prompting learners to explain underlying principles can influence transfer \cite{transFadeWrok}, while the appropriate level of guidance may vary with prior knowledge. The available descriptions of educational interventions provide limited evidence about how these elements are sequenced or adjusted to address learners' difficulties or prior experience levels.

This finding does not establish that adaptive support is necessary in every learning context, nor does it demonstrate the superiority or inferiority of any particular pedagogical strategy. Instead, it raises an empirical question: how do instructional sequencing, feedback, and scaffolding contribute to developing PQC knowledge and ensuring practical performance?

\begin{implicationbox}{Recommendation:}
Future studies should evaluate scaffolded learning pathways that progress from foundational concepts to independent implementation, integration, testing, and migration. Support mechanisms could include worked examples, guided exercises, contextual feedback, and debugging assistance. Evaluate adaptive or personalised support across learners with different prior competencies, focusing on knowledge, independent task completion, and security performance to determine which approaches best support PQC learning.
\end{implicationbox}

\subsection{G5: Security Focused Assessment Gap}

\textbf{Interpretation:} F5 identifies a gap between opportunities to engage in practical activities and evidence that learners can perform those activities securely. This distinction is consequential for PQC education because functional success does not necessarily establish secure implementation. An application may execute a cryptographic operation successfully while still containing weaknesses in algorithm or parameter selection, API usage, key handling, configuration, error handling, or integration \cite{acar2017cryptographic, toruan2026security, crptoImmerr, hekkala_implementing_2023}.

The available evidence therefore supports a more specific concern than a general shortage of assessment: practical performance and security correctness are not consistently documented as explicit assessment targets. Assessment formats such as quizzes, coding exercises, and projects can serve different purposes, but their presence alone does not establish whether learners' engineering work is evaluated against defined security requirements.

The limited reporting of security-related criteria also constrains the interpretation of evaluation outcomes. Even when a study reports knowledge gains or successful task completion, it may remain unclear whether learners can recognise and avoid implementation weaknesses. Likewise, the absence of security criteria in a course description does not prove that the course lacks such criteria; it means that the available evidence does not establish their use.

\begin{implicationbox}{Recommendation:}
Future evaluations should assess functional and security correctness. Functional assessment should examine cryptographic operations and interoperability, while security assessment should cover algorithm and parameter selection, secure API usage, key management, configuration, error handling, and implementation-specific vulnerabilities. Use clear criteria and observable evidence, including code, configurations, and test results. Researchers should report assessment tasks, scoring methods, and results transparently to support replication and evaluate readiness for secure PQC implementation and integration.
\end{implicationbox}

\subsection{G6: Long-Term Capability Validation Gap}

\textbf{Interpretation:} F6 identifies limited evidence that PQC-related learning persists over time and transfers to unfamiliar software engineering contexts. Immediate knowledge outcomes, successful completion of instructional exercises, and positive learner perceptions can provide useful evidence about learning, but they do not independently establish whether learners retain the required competencies or apply them in new situations \cite{nrc2000learning}.

This limitation matters because professional PQC adoption may require developers to work with unfamiliar libraries, different application architectures, existing cryptographic dependencies, or migration constraints that were not encountered during instruction. Performance on a practised exercise may not predict performance under these changed conditions. Evidence of retention and transfer is therefore needed to understand whether educational outcomes extend beyond the original learning environment.

The available evidence does not conclusively establish delayed practical capability or transfer under the review's criteria. This should be interpreted as a limitation of the evidence reported, rather than proof that learning is not retained or cannot transfer.

\begin{implicationbox}{Recommendation:}
Future evaluations should include delayed assessments of knowledge retention and transfer tasks to examine whether learners apply PQC skills in unfamiliar situations. Researchers should report assessment intervals, available assistance, task requirements, and evaluation criteria, distinguishing functional correctness from security performance. This would strengthen evidence of whether PQC competencies persist over time and transfer to new contexts.
\end{implicationbox}

\subsection{Implications for Software Engineering Education and Training}

The six identified gaps highlight opportunities to strengthen PQC education for software developers. However, these findings must be interpreted in light of the reporting limitations identified in the quality appraisal, as undocumented practices or outcomes cannot be assumed absent. Nevertheless, the findings highlight areas requiring greater transparency and empirical investigation, informing three priorities for future PQC education and research.

First, \textit{authentic and competency-aligned learning} should connect foundational knowledge with realistic engineering responsibilities. Tasks should be explicitly linked to the required knowledge and skills, learning activities, and observable assessment criteria to clarify the progression from conceptual understanding to applied capability.

Second, \textit{security-centred performance assessment} should distinguish task completion from secure implementation. Evaluations should examine functional correctness and security requirements using observable engineering artefacts. This would provide stronger evidence of learners' ability to implement and integrate PQC securely.

Third, \textit{empirical validation in professional contexts} should examine learning among practising developers and assess whether competencies persist over time and transfer to unfamiliar tasks. Although the reviewed training offerings' descriptions cover operationally relevant topics, they do not provide reported empirical evidence of learner evaluation.

These priorities constitute an evidence-informed research agenda rather than a validated educational framework. Future studies should operationalise and evaluate these practices to determine which combinations of authentic tasks, competency alignment, pedagogical support, and security-focused assessment best develop secure and transferable PQC engineering capability.

\section{Limitations}
\label{limitation}

Our study has several limitations. First, database coverage, search terms, and limited public documentation may have led us to miss relevant PQC educational interventions, particularly internal industry training. Second, distinguishing conceptual coverage from practical activities required interpretation, while our reviewer-derived NICE competency mappings indicate potential alignment rather than demonstrated learner competence. Third, despite predefined coding criteria and discussion to resolve disagreements, ambiguous source descriptions may have introduced subjectivity. Fourth, differences in purpose and reporting detail limited comparisons between research papers and training offerings; moreover, eight of the 15 training offerings lacked identifiable dates, limiting assessment of their currency. Finally, the small, heterogeneous sample and limited developer-focused evidence constrain generalisability. Our findings therefore reflect the verifiable evidence in the included sources and do not establish the completeness of existing provision or its effectiveness in developing practical competencies for secure PQC implementation and migration.

\section{Conclusion}
\label{Conclusion}

This multivocal literature review examined 23 sources (8 research papers and 15 training offerings) to investigate how PQC education prepares learners for adoption. All sources covered PQC foundations and algorithms, but practical software-engineering tasks received uneven attention, and explicitly developer-oriented education remained limited in the reviewed research. The findings highlight gaps between intended learning outcomes and demonstrated capabilities. Practical activities and assessments do not necessarily establish learners’ ability to implement and integrate PQC securely. Evidence of security-focused assessment, long-term retention, and transfer to unfamiliar tasks also remains limited. We identify six priorities for future research: evaluating education with practising developers; incorporating authentic engineering tasks; aligning tasks, competencies, and assessments; improving feedback and learner support; assessing functional and security correctness; and evaluating retention and transfer over time. This review provides an evidence map of PQC educational provision and an agenda for designing and evaluating developer-oriented education that supports secure, practical, and transferable PQC implementation and migration skills.

\bibliographystyle{ieeetr}
\bibliography{Sok_EduGapBib}

\appendix

\section*{Ethical Considerations}
\label{ethics}
This study reviewed previously published research works. No human participants were recruited, and no personal, sensitive, or identifiable data were collected or analysed. Accordingly, formal human-subjects ethical approval was not required.

\section*{Generative AI Use}
\label{AIuse}
The authors used generative AI tools to assist with manuscript revision, improve grammar, clarity, and readability, and support LaTeX formatting. The authors critically reviewed and verified all AI-assisted content and retain full responsibility for the final manuscript, including its methodology, analysis, findings, and conclusions. AI tools were not used for literature selection, evidence extraction, analysis, or research decision-making.

\end{document}